\documentclass{ws-procs9x6}

\newcommand{\hetrois}    {\mbox{$ ^{3}{\mathrm{He}}~                            $}}
\newcommand{\hetro}    {\mbox{$ ^{3}{\mathrm{He}^{+}}~                            $}}
\newcommand{\cobalt}    {\mbox{$ ^{57}{\mathrm{Co}}~                            $}}

\newcommand{\fluor}    {\mbox{$ ^{19}{\mathrm{F}}~                            $}}
\newcommand{\cfour}    {\mbox{$ {\mathrm{CF}_{4}}~                            $}}

\newcommand{\hequatre}    {\mbox{$ ^{4}{\mathrm{He}}~                            $}}
\newcommand{\neut}{$\tilde{\chi}^0$}
\newcommand{\neutt}{$\tilde{\chi}~$}
\newcommand{\mchi}{${\rm M_{\tilde{\chi}}}$}

\newcommand{\gams} {{$\gamma$-rays}}

\def\NIMA#1#2#3{{\rm Nucl.~Instr.~and~Meth.} {\bf{A#1}} (#2) #3}

\def\PLB{{\em Phys. Lett.}  B}

\def\PRL#1#2#3{{\rm Phys.~Rev.~Lett.} {\bf{#1}} (#2) #3}
\def\PLB#1#2#3{{\rm Phys.~Lett.} {\bf{B#1}} (#2) #3}

\def\APJ#1#2#3{{\rm Astrophys.~J.} {\bf{#1}} (#2) #3}
\def\APJS#1#2#3{{\rm Astrophys.~J.~Suppl.} {\bf{#1}} (#2) #3}
\def\AA#1#2#3{{\rm Astron. \& Astrophys.} {\bf{#1}} (#2) #3}
\def\JCAP#1#2#3{{\rm JCAP} {\bf{#1}} (#2) #3}

\begin{document}

\title{MIMAC-He3 : MIcro-tpc MAtrix of Chambers of \hetrois }

\author{D. Santos, O. Guillaudin, Th. Lamy, F. Mayet and E. Moulin}

\address{Laboratoire de Physique Subatomique et de Cosmologie,\\ 
 CNRS/IN2P3 et Universit\'e Joseph Fourier (Grenoble-1),\\ 
 53, avenue des Martyrs, 38026 Grenoble cedex, France
}

\begin{abstract}
The project of a micro-TPC matrix of chambers of \hetrois 
for direct detection of non-baryonic dark matter is 
outlined. The privileged properties of \hetrois  are highlighted. The double detection (ionization - projection of tracks) will assure  the electron-recoil discrimination. 
The complementarity of MIMAC-He3 for supersymmetric dark matter search with respect to other experiments is illustrated.The modular character of the detector allows to have different gases to get A-dependence. The pressure degreee of freedom gives the possibility to work at high and low pressure. The low pressure regime gives the possibility to get the directionality of the tracks. The first measurements of ionization at very few keVs for \hetrois in \hequatre gas are described. 

\end{abstract}


\bodymatter

\section{Introduction}\label{sec:intro}
Strong evidence in favor of the existence of non-baryonic dark matter arises from different cosmological observations.
The cosmic microwave backgroung (CMB) data ~\cite{wmap,archeops} in combination with high redshift supernov\ae~ analysis ~\cite{snap} 
and large scale structure surveys~\cite{sdss} seem to converge on an unified cosmological model~\cite{cosmomodel}. 
The non-baryonic cold dark matter (CDM) would consist of still not detected particles, among those being the generally referred to as 
WIMPs (Weakly Interacting Massive Particles) the privileged ones.
Among the different possible WIMPs, the lightest supersymmetric particle, in most scenarii, the lightest 
neutralino \neutt predicted by SUSY theories with R-parity conservation, stands as a well motivated 
candidate.\\ 
In the last decades, huge experimental efforts on a host of techniques in the field of direct search of
non-baryonic dark matter have been performed ~\cite{edelweiss,cdms,seidel,electronsmache3}.
Several detectors reached sufficient sensitivity to begin to test regions of the SUSY parameter space.
However, Wimp events have not yet been reported. Besides the fact that the cross section could be very weak, the energy threshold effect combined with the use of a heavy target nucleus 
leads to significant sensitivity loss for relatively light WIMPs ( 6 GeV $\leq ~$\mchi $\leq ~$40 GeV ).\\
As reported elsewhere~\cite{dm2000,nima,Santos:2005xj}, the use of
\hetrois~ as a target nucleus is motivated by its privileged features for dark matter search compared with other 
target nuclei.    
First, \hetrois~being a spin 1/2 nucleus, a detector made of such a material will be sensitive to the spin-dependent
interaction, leading to a natural complementarity to most existing or planned Dark Matter detectors 
($\nu$ telescopes, scalar direct detection as well as proton based spin-dependant detectors). 
In particular, it has been shown \cite{mache3plb,mimache3plb} that an \hetrois based detector will
present a good sensitivity  to low mass \neutt, within the framework of effective MSSM models
without gaugino mass unification at the GUT scale ~\cite{gondolo,belanger}.\\   
\begin{figure}
\begin{center}
\psfig{file=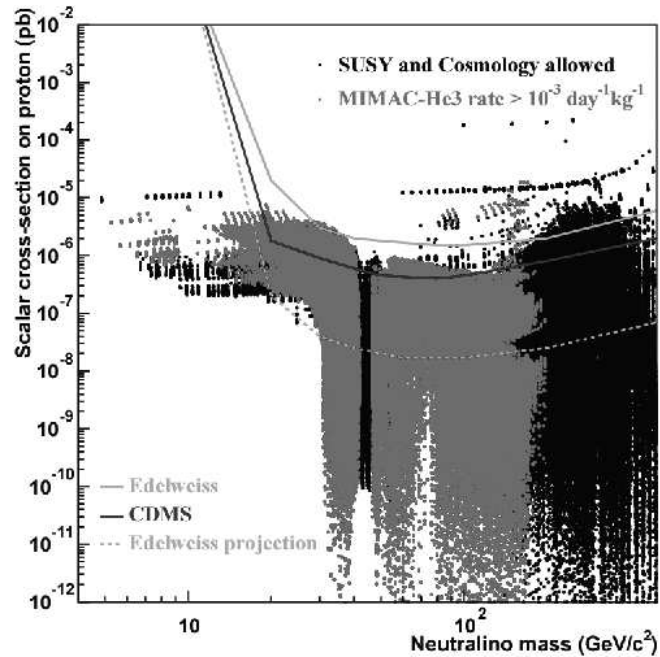,width=3.4in}
\end{center}
\caption{SUSY non minimal models, calculated with DarkSusy code ~\cite{gondolo}. In grey the models giving an axial cross section (\neutt - \hetrois) higher than the exclusion plot of MIMAC-\hetrois with 10kg  ~\cite{mimache3plb}. 
These models are compared with exclusion plots of scalar experiments and their projections. There are models, in grey, that will be very difficult to get with only a scalar approach.}
\label{aba:fig1}
\end{figure}

The \hetrois~presents in addition the following advantages with respect to other sensitive materials
for WIMPs detection :\\
- a very low Compton cross-section to gamma rays, two orders of magnitude weaker than in Ge : ${\rm 9\times10^{-1}}$ barns
for 10 keV \gams~ \\
- the neutron signature made possible by the capture process : $${\rm 
n\,+\,\hetrois\,\rightarrow\,p\,+^3H\,+\,764\,keV}$$ 
Indeed it allows for an easy discrimination with \neut signal ($\rm E \lesssim 6 \ keV$).
This property is a key point for Dark Matter search as neutrons in underground laboratories are considered as the
ultimate background.\\ 
Any dark matter detector should be able to separate a \neutt~event from the neutron background. 
Using energy measurement and electron-recoil discrimination,
MIMAC-He3 presents a high rejection for neutrons due to capture and multi-scattering of neutrons~\cite{thesemanu}.
The MIMAC project propose a modular detector in which different gases (\hetrois, \cfour) can be used to have a dependence on the mass of the target. The \fluor is other good target nucleus choice to have the axial interaction open, but proton based, increasing the attractiveness of the detector.

The MIMAC detector has two different regimes of work: i) high pressure (1,2 or 3 bar) and ii) low pressure (100 - 200 mbar). These two regimes allow us to have Wimp events at high pressure and search for correlation with the galactic halo apparent movement at low pressure. This last possibility should be validated with a special read out electronics as an important step of the project.



\noindent
\section{\bf {Micro-TPC and ionization-track projection detection}}\label{sec:mimac}

The micro time projection chambers with an avalanche amplification using a pixelized  
anode presents the required features to discriminate electron - recoil events with the 
double detection of the ionization energy and the track projection onto the anode.  
In order to get the electron-recoil discrimination, the pressure of the TPC should be 
such that the electron tracks with an energy less than 6 keV could be well resolved from 
the recoil ones at the same energy convoluted by the quenching factor.
The electrons produced by the primary interactions will drift to the amplification region (mesh) in a diffusion process following 
the well known distribution characterized by a radius of $\rm D\simeq \lambda \sqrt(L[cm]) $ where $\lambda $ is 
tipically 200 $\mu m $ for \hetrois at 1 bar and $ \rm L$ 
is the total drift in the chamber up to the mesh. This process has been simulated with Garfield 
and the drift velocities estimated  as a function of the pressure and the electric field. 
A typical value of $\rm 26 \mu  m/ns$ is obtained for $\rm 1\ kV/cm$ in pure \hetrois at a pressure of 1 bar. 
To prevent confusion between electron track projection and recoil ones the 
total drift length should be limited to L$\simeq $15 cm. It defines the 
elementary cell of the detector matrix and the simulations performed on the ranges of electrons 
and recoils suggest that with an anode of $\rm 350 \mu m$  the electron-recoil discrimination required can be obtained.
The quenching factor is an important point that should be addressed to quantify the 
amount of the total recoil energy recovered in the ionization channel. No measurements of the 
quenching factor (QF) in \hetrois have been  reported. However, an estimation can be obtained applying   the Lindhard calculations ~\cite{lindhard}.  
The estimated quenching factor given by Lindhard's theory for \hetrois shows up to 70 \% of the recoil energy 
going to the ionization channel for 5 keV  \hetrois recoil. 

\noindent
\section{{\bf Source MIMAC}}

In order to measure the QF for \hetrois and \hequatre we have developed at the LPSC a dedicated facility producing very
light ions at a few keV energies. This facility, called source MIMAC, incorporates an ECR ion source coupled to a Wien filter, selecting q/m,  and a high voltage extraction going up to 50 kV. 
\begin{figure}
\begin{center}
\psfig{file=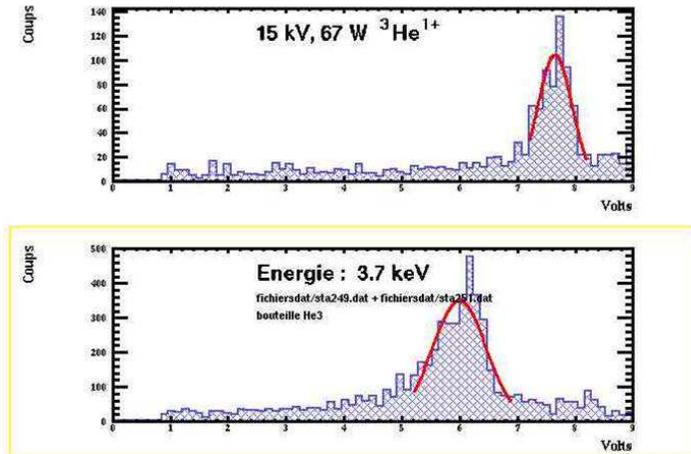,width=3.7in}
\end{center}
\caption{Time of flight measurements  performed with the MIMAC source. The figure shows the spectra at the two different positions (close to and far from the interface (source-chamber)) used to measure the \hetro ions output energy when they have been accelerated at 15kV}
\label{aba:fig2}
\end{figure}

The characterization of the output energies is made by a separate time of flight measurements as we can see on fig.2 for the case of \hetrois ions accelerated at 15 kV having a mean output energy of 3.7 keV.
Using this facility we can explore the ionization at very low energies for \hetrois ions. We have measured by TOF, five output energies going from 13.7 keV up to  3.7 keV corresponding to five values from 30 to 15 kV  of accelerating voltage extraction. Ionization measurements have been performed, with a standard micromegas grid in a gas chamber (95\% of \hequatre and 5\% of isobutane at 1 bar). A linear calibration fits very well the points measured and extrapolating to even lower voltage extraction, we can estimate the maximum output energy corresponding to 10.5 kV to 800 eV. On fig.3 the spectrum of the ionization left in the chamber by  \hetrois at 800 eV is shown. On the same spectrum we show an internal conversion electron spectrum of \cobalt during the two minutes the beam of \hetrois was on. This \cobalt source will allow us to get an idea of the equivalent electron energies. We can differentiate on the spectrum the peak of ionization well separated from the electronic noise. 
\begin{figure}
\begin{center}
\psfig{file=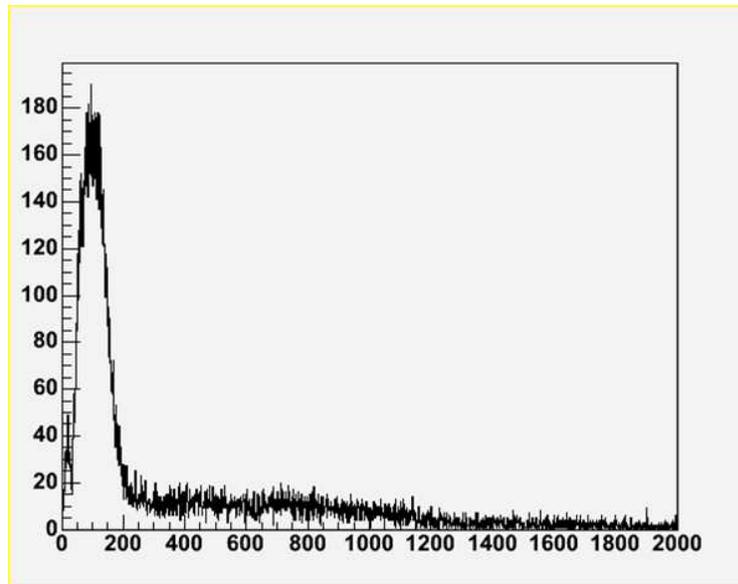,width=3.9in}
\end{center}
\caption{A two minutes spectrum showing ionization peak corresponding to a beam, produced by the MIMAC source, of \hetrois at an energy estimated to 800 eV. An internal conversion source of \cobalt spectrum is shown on the same spectrum. This source will help us to get the electron energy calibration for the QF measurement.}
\label{aba:fig3}
\end{figure}

\newpage

This spectrum shows clearly that we can expect to get the ionization left by \hetrois recoils in a chamber up to energies lower than 1 keV using the micromegas detector technology of our collaborators at Saclay ~\cite{ioannis}.

The next steps will be to perform the QF measurements with the setup described above and to show the track of \hetrois recoils with a dedicated read-out electronics that is building up at our laboratory.

\bibliography{ws-pro-sample}

\begin{thebibliography}{99}

\bibitem{archeops}A.~Beno\^{\i}t {\it et al.}, \AA{399}{2003}{L25}, M. Tristram {\it et al.}, \AA{436}{2005}{785}
\bibitem{wmap}D.~Spergel {\it et al.}, \APJS{148}{2003}{175}
\bibitem{snap}S.~Perlmutter {\it et al.}, \PRL{83}{1999}{670}
\bibitem{sdss}M.~Tegmark {\it et al.}, \APJ{606}{2004}{702}
\bibitem{cosmomodel} U.~Seljak {\it et al.} astro-ph/0604335
\bibitem{belanger}G. B\'elanger {\it et al.} hep-ph/0502079
\bibitem{gondolo}P.~Gondolo {\it et al.} \JCAP{0407}{2004}{008}
\bibitem{edelweiss}A.~Beno\^{\i}t {\it et al.}, \PLB{545}{2002}{43}
\bibitem{cdms}D.~Akerib {\it et al.}, \PRL{93}{2004}{211301}
\bibitem{seidel}W.~Seidel {\it et al.}, Proc.of the 4$^{\rm th}$ Intern. Conf. on Dark Matter in Astro and Particle Physics (DARK 2002), Feb. 2002, 
Cape Town (South Africa), Eds. H.-V. Klapdor-Kleingrothaus {\it et al.}, Springer, pp. 517
\bibitem{lindhard}J.~Lindhard {\it et al.}, Mat. Fys. Medd. K. Dan. Vidensk. Selsk. 33 (1963) 1-42.
\bibitem{electronsmache3}E.~Moulin {\it et al.}, \AA{453}{2006}{761}



\bibitem{dm2000}D.~Santos {\it et al.}, Proc. of the
${\textrm 4^{th}}$ Intern. Symposium on Sources 
and Detection of Dark Matter and Dark Energy in the Universe (DARK 2000), Feb. 2000, 
Marina Del Rey (CA, USA), Ed. D.B. Cline, Springer, pp. 469, 
\bibitem{ioannis}Y. Giommataris {\it et al.} \NIMA{376}{1996}{29}
\bibitem{nima}F.~Mayet {\it et al.}, \NIMA{455}{2000}{554}

\bibitem{Santos:2005xj}D.~Santos {\it et al.}, J.\ Phys.\ Conf.\ Ser.\  {\bf 39} (2006) 154

 




\bibitem{mache3plb}F.~Mayet {\it et al.}, \PLB{538}{2002}{257}

\bibitem{mimache3plb}E.~Moulin {\it et al.}, \PLB{614}{2005}{143}

\bibitem{mimache3}D.~Santos, F.~Mayet {\it et al.}, {\it in preparation}
\bibitem{thesemanu}E.~Moulin, PhD Thesis, 2005, Universit\'e J. Fourier, Grenoble, France
 
  
  

 \end{thebibliography}


\end{document}